\documentclass[twocolumn,showpacs,preprintnumbers,superscriptaddress,amsmath,amssymb,prl]{revtex4}

\usepackage{graphicx}% Include figure files
\usepackage{dcolumn}% Align table columns on decimal point
\usepackage{bm}% bold math
\usepackage{epsfig}% bold math
\newcommand \be{\begin{equation}}
\newcommand \ba{\begin{eqnarray}}
\newcommand \ea{\end{eqnarray}}
\newcommand \ee{\end{equation}}

\begin{document}

\preprint{Inverse statistics in turbulence}

\title{Inversion formula of multifractal energy dissipation in 3D
fully developed turbulence}

\author{Jian-Liang Xu}
\affiliation{Institute of Clean Coal Technology, East China
University of Science and Technology, Box 272, Shanghai, 200237,
People's Republic of China}
\author{Wei-Xing Zhou}
\email{wxzhou@moho.ess.ucla.edu}
\affiliation{School of Science and
School of Business, East China University of Science and Technology,
Shanghai 200237, China}
\author{Hai-Feng Liu}
\author{Xin Gong}
\author{Fu-Cheng Wang}
\author{Zun-Hong Yu}
\affiliation{Institute of Clean Coal Technology, East China
University of Science and Technology, Box 272, Shanghai, 200237,
People's Republic of China}

\date{\today}

\begin{abstract}
The concept of inverse statistics in turbulence has attracted much
attention in the recent years. It is argued that the scaling
exponents of the direct structure functions and the inverse
structure functions satisfy an inversion formula. This proposition
has already been verified by numerical data using the shell model.
However, no direct evidence was reported for experimental three
dimensional turbulence. We propose to test the inversion formula
using experimental data of three dimensional fully developed
turbulence by considering the energy dissipation rates in stead of
the usual efforts on the structure functions. The moments of the
exit distances are shown to exhibit nice multifractality. The
inversion formula between the direct and inverse exponents is then
verified.
\end{abstract}


\maketitle

%
%\section{Introduction}
%\label{sec:intro}

Intermittency of fully developed isotropic turbulence is well
captured by highly nontrivial scaling laws in structure functions
and multifractal nature of energy dissipation rates
\cite{Frisch-1996}. The {\it{direct}} (longitudinal) structure
function of order $q$ is defined by $S_q({r})\equiv \langle
\Delta{v}_{\parallel}({r})^q\rangle$, where
$\Delta{v}_{\parallel}({r})$ is the longitudinal velocity difference
of two positions with a separation of $r$. The anomalous scaling
properties characterized by $S_q({r})\sim r^{\zeta(q)}$ with a
nonlinear scaling exponent function $\zeta(q)$ were uncovered
experimentally \cite{Anselmet-Gagne-Hopfinger-Antonia-1984-JFM}.

While the direct structure functions consider the statistical
moments of the velocity increments $\Delta{v}$ measured over a
distance $r$, the inverse structure functions are concerned with the
exit distance $r$ where the velocity fluctuation exceeds the
threshold $\Delta{v}$ at minimal distance \cite{Jensen-1999-PRL}. An
alternative quantity is thus introduced, denoted the distance
structure functions \cite{Jensen-1999-PRL} or inverse structure
functions
\cite{Biferale-Cencini-Vergni-Vulpiani-1999-PRE,Biferale-Cencini-Lanotte-Vergni-Vulpiani-2001-PRL},
that is, $T_q(\Delta v) \equiv \langle{{r}^q(\Delta v)}\rangle
\label{Eq:ISF1}$. Due to the duality between the two methodologies,
one can intuitively expected that there is a power-law scaling
stating that $T_p({\Delta{v}})\sim {\Delta{v}}^{\phi(p)}$, where
$\phi(p)$ is a nonlinear concave function. This point is verified by
the synthetic data from the GOY shell model of turbulence exhibiting
perfect scaling dependence of the inverse structure functions on the
velocity threshold \cite{Jensen-1999-PRL}. Although the inverse
structure functions of two-dimensional turbulence exhibit sound
multifractal nature
\cite{Biferale-Cencini-Lanotte-Vergni-Vulpiani-2001-PRL}, a
completely different result was obtained for three-dimensional
turbulence, where an experimental signal at high Reynolds number was
analyzed and no clear power law scaling was found in the exit
distance structure functions
\cite{Biferale-Cencini-Vergni-Vulpiani-1999-PRE}. Instead, different
experiments show that the inverse structure functions of
three-dimensional turbulence exhibit clear extended self-similarity
\cite{Beaulac-Mydlarski-2004-PF,Pearson-vandeWater-2005-PRE,Zhou-Sornette-Yuan-2006-PD}.

For the classical binomial measures, Roux and Jensen
\cite{Roux-Jensen-2004-PRE} have proved an exact relation between
the direct and inverse scaling exponents,
\begin{equation}
\left\{
\begin{array}{lll}
\zeta(q) &=& -p\\
\phi(p) &=& -q
\end{array}
\right.~, \label{Eq:ZetaTheta}
\end{equation}
which is verified by the simulated velocity fluctuations from the
shell model. This same relation is also derived intuitively in an
alternative way for velocity fields \cite{Schmitt-2005-PLA}. A
similar derivation was given for Laplace random walks as well
\cite{Hastings-2002-PRL}. However, this prediction
(\ref{Eq:ZetaTheta}) is not confirmed by wind-tunnel turbulence
experiments (Reynolds numbers ${\mbox{Re}}=400-1100$)
\cite{Pearson-vandeWater-2005-PRE}. We argue that this dilemma comes
from the ignoring of the fact that velocity fluctuation is not a
conservative measure like the binomial measure. In other words,
Eq.~(\ref{Eq:ZetaTheta}) can not be applied to nonconservative
multifractal measures.

Actually, Eq.~(\ref{Eq:ZetaTheta}) is known as the inversion formula
and has been proved mathematically for both discontinuous and
continuous multifractal measures
\cite{Mandelbrot-Riedi-1997-AAM,Riedi-Mandelbrot-1997-AAM}. Let
$\mu$ be a probability measure on $[0,1]$ with its integral function
$M(t) = \mu([0,t])$. Then its inverse measure can be defined by
\begin{equation}
\mu^\dag = M^\dag(s) = \left\{
\begin{array}{lll}
\inf\{t:M(t)>s\}, && {\mbox{if }} s<1\\
1, && {\mbox{if }} s=1
\end{array}
\right.~, \label{Eq:invMu}
\end{equation}
where $M^\dag(s)$ is the inverse function of $M(t)$. If $\mu$ is
self-similar, then the relation $\mu = \sum_{i=1}^n
p_i\mu(m_i^{-1}(\cdot))$ holds, where $m_i$'s are similarity maps
with scale contraction ratios $r_i\in(0,1)$ and $\sum_{i=1}^np_i=1$
with $p_i>0$. The multifractal spectrum of measure $\mu$ is the
Legendre transform $f(\alpha)$ of $\tau$, which is defined by
\begin{equation}
\sum_{i=1}^n p_i^qr_i^{-\tau}=1~. \label{Eq:tau}
\end{equation}
It can be shown that
\cite{Mandelbrot-Riedi-1997-AAM,Riedi-Mandelbrot-1997-AAM}, the
inverse measure $\mu^\dag$ is also self-similar with ratio
$r_i^\dag=p_i$ and $p_i^\dag=r_i$, whose multifractal spectrum
$f^\dag(\alpha^\dag)$ is the Legendre transform of $\theta$, which
is defined implicitly by
\begin{equation}
\sum_{i=1}^n (p_i^\dag)^p (r_i^\dag)^{-\theta}=1~. \label{Eq:theta}
\end{equation}
It is easy to verify that the inversion formula holds that
\begin{equation}
\left\{
\begin{array}{lll}
\tau(q) &=& -p\\
\theta(p) &=& -q
\end{array}
\right.~. \label{Eq:TauTheta}
\end{equation}
Two equivalent testable formulae follow immediately that
\begin{equation}\label{Eq:InvForm1}
\tau(q) = -\theta^{-1}(-q)
\end{equation}
and
\begin{equation}\label{Eq:InvForm2}
\theta(p) = -\tau^{-1}(-p)~.
\end{equation}
Due to the conservation nature of the measure and its inverse in the
formulation outlined above, we figure that it is better to test the
inverse formula in turbulence by considering the energy dissipation.

Very good quality high-Reynolds turbulence data have been collected
at the S1 ONERA wind tunnel by the Grenoble group from LEGI
\cite{Anselmet-Gagne-Hopfinger-Antonia-1984-JFM}. We use the
longitudinal velocity data obtained from this group. The size of the
velocity time series that we analyzed is $N \approx 1.73 \times
10^7$.

The mean velocity of the flow is approximately $\langle{v}\rangle =
20 $m/s (compressive effects are thus negligible). The
root-mean-square velocity fluctuations is $v_{\mathtt{rms}} = 1.7
$m/s, leading to a turbulence intensity equal to $I =
{v_{\mathtt{rms}}} / {\langle{v}\rangle} = 0.0826$. This is
sufficiently small to allow for the use of Taylor's frozen flow
hypothesis. The integral scale is approximately $4 \mathtt{m}$ but
is difficult to estimate precisely as the turbulent flow is neither
isotropic nor homogeneous at these large scales.

The Kolmogorov microscale $\eta$ is given by
\cite{Meneveau-Sreenivasan-1991-JFM} $\eta = \left[\frac{\nu^2
\langle{v}\rangle^2}{15 \langle(\partial v/\partial t)^2\rangle
}\right]^{1/4} = 0.195 \mathtt{mm}$, where $\nu = 1.5 \times 10^{-5}
\mathtt{m^2 s^{-1}}$ is the kinematic viscosity of air. $\partial
v/\partial t$ is evaluated by its discrete approximation with a time
step increment $\partial t = 3.5466 \times 10^{-5} \mathtt{s}$
corresponding to the spatial resolution $\epsilon = 0.72
\mathtt{mm}$ divided by $\langle{v}\rangle$, which is used to
transform the data from time to space applying Taylor's frozen flow
hypothesis.

The Taylor scale is given by \cite{Meneveau-Sreenivasan-1991-JFM}
$\lambda =\frac{\langle{v}\rangle v_{\mathtt{rms}}}{\langle
(\partial v/\partial t)^2 \rangle^{1/2}} =16.6 \mathtt{mm}$. The
Taylor scale is thus about $85$ times the Kolmogorov scale. The
Taylor-scale Reynolds number is $Re_\lambda =
\frac{v_{\mathtt{rms}}\lambda}{\nu} = 2000$. This number is
actually not constant along the whole data set and fluctuates by
about $20\%$.

We have checked that the standard scaling laws
previously reported in the literature are recovered with this
time series. In particular, we have verified the validity
of the power-law scaling $E(k) \sim k^{-\beta}$ with an exponent
$\beta$ very close to $\frac{5}{3}$ over a range more than two
decades, similar to Fig. 5.4 of \cite{Frisch-1996} provided by
Gagne and Marchand on a similar data set from the same
experimental group. Similarly, we have checked carefully the
determination of the inertial range by combining the scaling
ranges of several velocity structure functions (see Fig. 8.6 of
\cite[Fig. 8.6]{Frisch-1996}). Conservatively, we are led to a
well-defined inertial range $60 \leq {r}/\eta \leq 2000$.

The exit distance sequence $r(\delta{E})=\{r_j(\delta{E})\}$ for a
given energy threshold $\delta E$ can be obtained as follows. For a
velocity time series $\{v_i=v(t_i):i=1,2,\cdots\}$, the energy
dissipation rate series is constructed as $\{E_i=(v_{i+1}-v_i)^2\}$.
We assume that $E_i$ is distributed uniformly on the interval
$[t_i,t_{i+1})$. A right continuous energy density function is
constructed such that $e(t) = E_i ~{\rm{for}}~ t\in [t_i,t_{i+1})$.
The exit distance sequence $\{r_j(\delta{E})\}$ is determined
successively by $\sum_{k=1}^j {r_k}/{\langle{v}\rangle} = \inf
\{t:\int_0^t e(t)dt \geqslant j \cdot\delta{E}\}$. Since energy is
conservative, we have
\begin{equation}
r_j(2\delta{E})=r_{2j-1}(\delta{E})+r_{2j}(\delta{E})~.
\label{Eq:Additive}
\end{equation}
With this relation, we can reduce the computational time
significantly. In order to determine $r_i(\delta{E})$, we choose a
minimal threshold $E_{\min}$, one tenth of the mean of $\{E_i\}$,
and obtain $r_i(E_{\min})$. Then other sequences of $r_i$ for
integer $\delta{E}/E_{\min}$ can be easily determined with relation
(\ref{Eq:Additive}).

\begin{figure}[!h]
\begin{center}
\includegraphics[width=8cm]{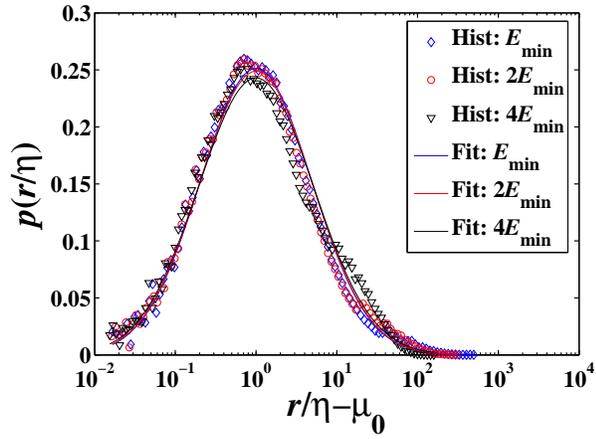}
\caption{(Color online) Empirical probability density functions of
exit distance $r/\eta$ for energy increments $\delta{E} = E_{\min}$,
$2E_{\min}$, and $4E_{\min}$. The value of $\mu_0$ is the fitted
parameter of $\mu$ in the Gaussian distribution.} \label{Fig:PDF}
\end{center}
\end{figure}

In Fig.~\ref{Fig:PDF} is shown the empirical probability density
functions (pdf's) of exit distance $r/\eta$ for energy increments
$E_{\min}$, $2E_{\min}$, and $4E_{\min}$. At a fist glance, the
probability density functions are roughly Gaussian, as shown by the
continuous curves in Fig.~\ref{Fig:PDF}. The value of $\mu_0$ is the
fitted parameter of the mean $\mu$ in the Gaussian distribution. For
$r/\eta < \mu_0$, the three empirical pdf's collapse to a single
curve. However, for large $r/\eta > \mu_0$, the three empirical
pdf's differ from each other, especially in the right-hand-side tail
distributions. This discrepancy is the cause of the occurrence of
multifractal behavior of exit distance, which we shall show below.

An intriguing feature in the empirical pdf is emergence of small
peaks observed at $r/\epsilon =1,2,\cdots$ in the tail
distributions. Comparably, the pdf of exit distance of multinomial
measure exhibits clear singular peaks. Therefore, these small peaks
in Fig.~\ref{Fig:PDF} can be interpreted as finite-size truncations
of singular distributions, showing the underlying singularity of the
dissipation energy, which is consistent with the multifractal nature
of the exit distance of dissipation energy.

\begin{figure}[!htb]
\begin{center}
\includegraphics[width=8.5cm]{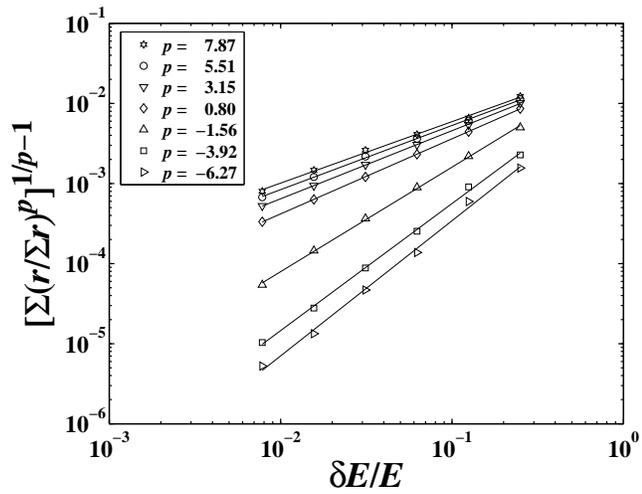}
\caption{Double logarithmic plots of $[T_p(r/\sum{r})]^{1/(p-1)}$
versus $\delta{E}/E$ for different values of $p$. The straight lines
are best fits to the data.} \label{Fig:Mr:dE}
\end{center}
\end{figure}

According to the empirical probability density functions, the
moments of exit distance exist for both positive and negative
orders. Figure \ref{Fig:Mr:dE} illustrates the double logarithmic
plots of $[T_p(r/\sum{r})]^{1/(p-1)}$ versus $\delta{E}/E$ for
different values of $p$. For all values of $p$, the power-law
dependence is evident. The straight lines are best fits to the data,
whose slopes are estimates of $\theta(p)/(p-1)$.

\begin{figure}[!htb]
\begin{center}
\includegraphics[width=8cm]{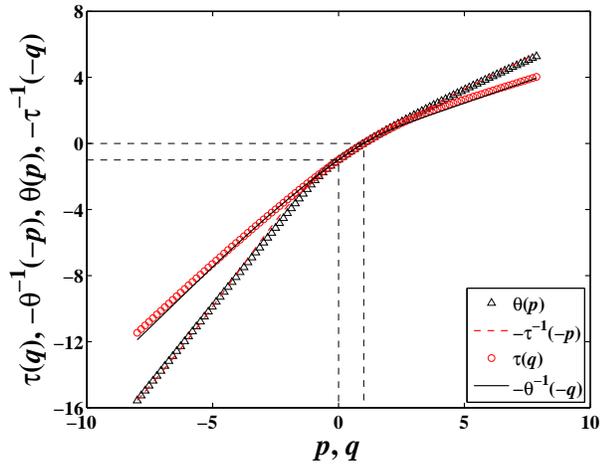}
\caption{(Color online) Testing the inversion formula of turbulence
dissipation energy.} \label{Fig:matching}
\end{center}
\end{figure}

The inverse scaling exponent $\theta(p)$ is plotted as triangles in
Fig.~\ref{Fig:matching} against order $p$, while the direct scaling
exponent $\tau(q)$ is shown as open circles. We can obtain the
function $-\tau^{-1}(-p)$ numerically from the $\tau(q)$ curve,
which is shown as a dashed line. One can observe that the two
functions $\theta(p)$ and $-\tau^{-1}(-p)$ coincide remarkably,
which verifies the inverse formulation (\ref{Eq:InvForm1}).
Similarly, we obtained $-\theta^{-1}(-q)$ numerically from the
$\theta(p)$ curve, shown as a solid line. Again, a nice agreements
between $\tau(q)$ and $-\theta^{-1}(-q)$ is observed, which verifies
(\ref{Eq:InvForm2}).

In summary, we have suggested to test the inversion formula in three
dimensional fully developed turbulence by considering the energy
dissipation rates in stead of the usual efforts on the structure
functions. The moments of the exit distances exhibit nice
multifractality. We have verified the inversion formula between the
direct and inverse exponents.

\begin{acknowledgements}
The experimental turbulence data obtained at ONERA Modane were
kindly provided by Y. Gagne. We are grateful to J. Delour and J.-F.
Muzy for help in pre-processing these data. This work was partly
supported by the National Basic Research Program of China (No.
2004CB217703) and the Project Sponsored by the Scientific Research
Foundation for the Returned Overseas Chinese Scholars, State
Education Ministry.
\end{acknowledgements}

\bibliography{Bibliography}

\end{document}